\documentclass[11pt]{article}
\makeatletter\if@twocolumn\PassOptionsToPackage{switch}{lineno}\else\fi\makeatother

      \makeatletter
\usepackage{wrapfig}
\newcounter{aubio}

\long\def\bioItem{\@ifnextchar[{\@bioItem}{\@@bioItem}}

\long\def\@bioItem[#1]#2#3{
 \stepcounter{aubio}
 \expandafter\gdef\csname authorImage\theaubio\endcsname{#1}
 \expandafter\gdef\csname authorName\theaubio\endcsname{#2}
 \expandafter\gdef\csname authorDetails\theaubio\endcsname{#3}
}

\long\def\@@bioItem#1#2{
 \stepcounter{aubio}
 \expandafter\gdef\csname authorName\theaubio\endcsname{#1}
 \expandafter\gdef\csname authorDetails\theaubio\endcsname{#2}
}

\def\printBio{\@tempcnta=0
   \loop
     \advance \@tempcnta by 1
     \def\aubioCnt{\the\@tempcnta}
     \setlength{\intextsep}{0pt}\setlength{\columnsep}{10pt}\newbox\boxa \setbox\boxa\vbox{\csname authorDetails\aubioCnt\endcsname}
     {\parindent0pt\textbf{\csname authorName\aubioCnt\endcsname}\csname authorDetails\aubioCnt\endcsname \par\bigskip}\ifnum\@tempcnta < \theaubio
   \repeat
   }

\makeatother

\usepackage{amsfonts,amssymb,amsbsy,latexsym,amsmath,tabulary,graphicx,times,xcolor}
\usepackage[utf8]{inputenc}
\usepackage{fancyhdr}
\def\NormalBaseline{\def\baselinestretch{1.1}}
\makeatletter
\def\hlinewd#1{\noalign{\ifnum0=`}\fi\hrule \@height #1\futurelet\reserved@a\@xhline}

\def\@shorttitle{\@empty}
\def\shorttitle#1{\gdef\@shorttitle{#1}}

\fancypagestyle{custom}{
\fancyhf{}
\fancyhead[C]{\@shorttitle}
\fancyhead[R]{\thepage}
\fancyfoot[C]{}

}
\fancypagestyle{plain}{
\fancyhf{}

}
\makeatother

\usepackage{times}
\usepackage[a4paper,margin=2.5cm,headsep=.7cm,headheight=18pt,top=3cm,footnotesep=1.5\baselineskip]{geometry}
\usepackage{caption}

\usepackage{url,multirow,morefloats,floatflt,cancel}
\makeatletter
\AtBeginDocument{\@ifpackageloaded{textcomp}{}{\usepackage{textcomp}}}
\makeatother
\usepackage{colortbl}
\usepackage{xcolor}
\usepackage{pifont}
\usepackage[nointegrals]{wasysym}
\makeatletter

\def\mcWidth#1{\csname TY@F#1\endcsname+\tabcolsep}
\def\cAlignHack{\rightskip\@flushglue\leftskip\@flushglue\parindent\z@\parfillskip\z@skip}
\def\rAlignHack{\rightskip\z@skip\leftskip\@flushglue \parindent\z@\parfillskip\z@skip}
\@ifundefined{etal}{}{}

\usepackage{ifxetex}
\ifxetex\else\if@twocolumn\@ifpackageloaded{stfloats}{}{\usepackage{dblfloatfix}}\fi\fi

\AtBeginDocument{
\expandafter\ifx\csname eqalign\endcsname\relax
\def\eqalign#1{\null\vcenter{\def\\{\cr}\openup\jot\m@th
  \ialign{\strut$\displaystyle{##}$\hfil&$\displaystyle{{}##}$\hfil
      \crcr#1\crcr}}\,}
\fi
}

\AtBeginDocument{\@ifpackageloaded{endfloat}{\renewcommand\efloat@iwrite[1]{\immediate\expandafter\protected@write\csname efloat@post#1\endcsname{}}}{\newif\ifefloat@tables}}
\def\BreakURLText#1{\@tfor\brk@tempa:=#1\do{\brk@tempa\hskip0pt}}
\let\lt=<
\let\gt=>
\def\processVert{\ifmmode|\else\textbar\fi}

\@ifundefined{subparagraph}{
\def\subparagraph{\@startsection{paragraph}{5}{2\parindent}{0ex plus 0.1ex minus 0.1ex}{0ex}{\normalfont\small\itshape}}}{}

\newcommand\role[1]{\unskip}
\newcommand\aucollab[1]{\unskip}

\@ifundefined{tsGraphicsScaleX}{}{}
\@ifundefined{tsGraphicsScaleY}{}{}
\def\fixFloatSize#1{}
\DeclareMathAlphabet{\mathpzc}{OT1}{pzc}{m}{it}
\def\URL#1#2{\@ifundefined{href}{#2}{\href{#1}{#2}}}
\def\UrlOrds{\do\*\do\-\do\~\do\'\do\"\do\-}\g@addto@macro{\UrlBreaks}{\UrlOrds}

\edef\fntEncoding{\f@encoding}

\makeatother

\newif\ifmultipleabstract\multipleabstractfalse 
\usepackage[numbers]{natbib}
\usepackage{titlesec}
\usepackage[T1]{fontenc}
\usepackage[
  hidelinks,
  pdftitle={Deterministic Narrative Gravity: Bridging Theory of Mind and Historical Simulation through ABM-LLM Architectures},
  pdfauthor={Alicia Colmenero-Fernandez, Lidia Ortega Alvarado},
  pdfsubject={To appear in the Journal of Computer Applications in Archaeology},
  pdfkeywords={Agent-Based Modeling, Large Language Models, deterministic simulation, narrative dialogue, cognitive architecture, multi-agent systems, historical simulation, Theory of Mind}
]{hyperref}

\titleformat{\section}[hang]{\NormalBaseline\filright\large\bfseries}{\large\thesection}{10pt}{}[]
\titleformat{\subsection}[hang]{\NormalBaseline\filright\bfseries}{\thesubsection}{10pt}{}[]
\titleformat{\subsubsection}[hang]{\NormalBaseline\filright\bfseries\itshape}{\upshape\thesubsubsection}{10pt}{}[]
\titleformat{\paragraph}[runin]{\NormalBaseline\filright\bfseries}{\theparagraph}{10pt}{}[]
\titleformat{\subparagraph}[runin]{\NormalBaseline\filright\bfseries\itshape}{\thesubparagraph}{10pt}{}[]

\titlespacing{\section}{0pt}{1.5\baselineskip}{.2\baselineskip}
\titlespacing{\subsection}{0pt}{1.5\baselineskip}{.2\baselineskip}
\titlespacing{\subsubsection}{0pt}{1.5\baselineskip}{.2\baselineskip}
\titlespacing{\paragraph}{0pt}{.5\baselineskip}{10pt}
\titlespacing{\subparagraph}{0pt}{.5\baselineskip}{10pt}

\begin{document}

\renewcommand*\rmdefault{bch}\normalfont\upshape

\shorttitle{}
\date{}

\title{\NormalBaseline\raggedright\bfseries Deterministic Narrative Gravity: Bridging Theory of Mind and Historical Simulation through ABM-LLM Architectures}

\def\AuAffLabelStyle#1{\textsuperscript{\upshape#1}}
\def\AuFont{\bfseries\large}
\def\AuSep{, }
\def\AffSep{\\}
\let\origthanks\thanks
\renewcommand\thanks[1]{\begingroup\let\rlap\relax\origthanks{#1}\endgroup}
\author{\hskip2pc\parbox{.82\linewidth}{\AuFont Alicia Colmenero-Fern\'{a}ndez\AuAffLabelStyle{1,2}\thanks{E-mail: acf00053@red.ujaen.es; lidia@ujaen.es}
      \AuSep \AuFont Lidia Ortega Alvarado\AuAffLabelStyle{1,3}
      \\[3pt]
    \normalfont\itshape\NormalBaseline \AuAffLabelStyle{1}
    University of Ja{\'e}n\normalfont\itshape\NormalBaseline \AffSep
    \normalfont\itshape\NormalBaseline \AuAffLabelStyle{2}
    IA-ismo LAB\unskip,\normalfont\itshape\NormalBaseline \AffSep
    \normalfont\itshape\NormalBaseline \AuAffLabelStyle{3}
    Departamento de Inform\'{a}tica\unskip, \normalfont\itshape\NormalBaseline }}

\maketitle
\pagestyle{custom}

\vspace{-2\baselineskip}
\begin{center}
  \small To appear in the Journal of Computer Applications in Archaeology
\end{center}
\vspace{-1\baselineskip}

\begin{abstract}\small\linespread{1.1}\selectfont
Historical social simulation faces a fundamental reproducibility challenge: the stochastic nature of large language models (LLMs) introduces a digital butterfly effect that prevents exact replication across multi-turn runs. Furthermore, the inability to control low-level server parameters in commercial LLM hosting environments severely limits the reproducibility required to systematically evaluate agent responses. To overcome this opacity, we present a deterministic multi-agent framework coupling LLMs to an Agent-Based Modelling (ABM) pipeline---grounded in Theory of Mind (ToM) and Cognitive Archaeology---and validate exact artifact identity within local, pinned execution environments using SHA-256 hashing. Run locally at temperature zero with clean per-call model reloads, the system injects differentiated cognitive states per turn through dissonance-driven behavioural modes and gated internal voices. Evaluating seven model configurations spanning 4B to 31B parameters against an ablated baseline reveals clear operational boundaries. Small models ($\le$8B) suffer from context capacity overload, producing repetitive, specular dialogue loops. However, at $\ge$12B parameters, the ABM scaffolding breaks the monotonous response patterns seen in ablation. Crucially, these scaffolded models exhibit spontaneous diegetic expansion: they coherently expand scenario lore through tactically motivated, anachronism-free details rather than escalating into aggression. By resisting the structural pull toward premature scene closure, scaffolded sessions sustain an open terminal Narrative Vector ($V_N = \text{Open}$ in 12/21 overall; 9/9 in official $\ge$12B configurations) whereas unscaffolded ablation never does (0/21). This framework provides a transparent, auditable benchmark architecture for virtual archaeology, with the full pipeline released as open-source software to enable independent replication.
\end{abstract}


\clearpage

\section{Introduction}

Agent-Based Models (ABM) constitute the dominant paradigm in social simulation \cite{ref1,ref2}; however, their deductive rigidity precludes the capture of human discursive agency. While contemporary architectures such as Generative Agents \cite{ref3}, WarAgent \cite{ref4}, and BattleAgent \cite{ref5} leverage Large Language Models (LLMs) to achieve greater naturalism, they inherit two critical vulnerabilities: stochasticity (lack of reproducibility) and alignment bias (sycophancy induced by RLHF). In the context of virtual archaeology, these vulnerabilities are particularly disqualifying: an unauditable simulation of past social interaction cannot function as scientific evidence, regardless of its narrative plausibility.

To address this, we present a hybrid ABM-LLM architecture grounded in controlled inference and transduction. We implement a Python engine that computes cognitive signals and dynamic anchors to counter what we term \emph{narrative gravity}. Hash comparison showed bit-exact repeatability in specified pinned environments, while cross-architecture tests exposed prompt-dependent numerical branching. The framework also provides exploratory evidence of coherent narrative continuation beyond immediate prompt resolution. It integrates Theory of Mind (ToM) dynamics---a construct central to Cognitive Archaeology's account of past intentionality---prioritizing the individual as the base unit for scaling collective dynamics and ensuring observability under the ODD (Overview, Design concepts, and Details) protocol.

\subsection{Reproducibility and Traceability as the Foundation of Scientific Method}

Historical research requires reproducible experiments, establishing experimental reproducibility as the minimum standard of scientific credibility \cite{ref6,ref7}. Without traceability, ABMs risk being narratively convincing yet scientifically opaque. As demonstrated by failed replications \cite{ref8}, omitting sampling mechanisms mandated by the original ODD standard invalidates replication. Consequently, our framework emphasizes explicability, reproducibility, and open-access code to ensure legitimate scientific auditing.

\subsection{State of the Art}

\subsubsection{The ABM Paradigm: Achievements and Limitations}

Since the foundational works of Epstein, Axtell, and others, ABMs have consistently demonstrated how local rules generate macroscopic social complexity \cite{ref1,ref9,ref10,ref11}. To ensure traceability, the field adopted the ODD protocol \cite{ref12,ref13,ref2}, facilitating transdisciplinary integration with Individual-Based Modeling (IBM) \cite{ref14}, although algorithmic transparency remains a persistent problem \cite{ref15,ref2}.

Classical social simulation mathematically models opinion dynamics through four interaction forces \cite{ref16}:

\begin{itemize}
  \item \textbf{Assimilation:} Attraction toward similar stances, with costs proportional to divergence.
  \item \textbf{Negotiation:} Consensus through proximity at a lower energetic cost.
  \item \textbf{Opposition:} Repulsion and polarization when messages cross a rejection threshold.
  \item \textbf{Reinforcement:} Validation and social contagion when messages align with existing beliefs.
\end{itemize}

Despite these frameworks, traditional ABMs face practical barriers: high computational costs relative to biological or social systems \cite{ref17,ref18,ref19}, and ``narrative reductionism,'' where hard-coded rules execute the \emph{what} but fail to generate the subjective \emph{why} \cite{ref18}. In fields like virtual archaeology, ABMs evaluate micro-macro causality and environmental affordances \cite{ref20,ref5,ref21}. To overcome algorithmic rigidity, our work adopts the ABM mathematical approach to interaction but uses LLMs to generate perception variables, avoiding narrative reductionism without the computational burden of massive multi-agent societies.

\subsection{LLMs as Infrastructure for Discursive Agency}

Social causality is fundamentally semantic, driven by interpretive frameworks and collective memories \cite{ref22,ref23,ref24,ref25}. Regardless of debates regarding their structural comprehension \cite{ref27} or stochastic nature \cite{ref26}, LLMs operate on the exact substrate of historical agency: contextualized language. LLM-based architectures produce complex emergent behaviors without explicit rules, outperforming traditional systems \cite{ref28} in simulating polarization and historical shocks \cite{ref3,ref29,ref30,ref31,ref32}.

However, decentralized autonomy introduces vulnerabilities like unguided drift, role flipping, and infinite loops \cite{ref33}, which provoke narrative collapse. Furthermore, multi-agent LLM systems frequently exhibit conformity biases and groupthink \cite{ref34}. A significant gap persists: the literature lacks approaches that organically integrate generative flexibility with rigorous structural control. Our work addresses this through fixed greedy-decoding, clean-reload controls, and a ToM-based parametric framework for navigating narrative gravity.

\subsection{Determinism and Inference in LLM-Based Social Simulation}

The inherent stochasticity of LLMs---arising from probabilistic token selection \cite{ref35} and hardware-level non-determinism \cite{ref36}---creates a digital ``butterfly effect'' that precludes replicability in long-horizon simulations \cite{ref26,ref37}. Current stabilization solutions remain insufficient: statistical averaging or interactive probing oversimplify complex interactions \cite{ref5,ref38,ref39,ref40,ref4,ref41}, while qualitative metrics \cite{ref34,ref42} and infrastructure-level fixes \cite{ref43,ref44} demand prohibitive supercomputing resources.

We address this through a local hybrid architecture using Ollama \cite{ref45}, fixed greedy-decoding settings, server-level Flash Attention disabling, and clean model reloads between calls. These controls produced bit-exact repetition within the tested pinned environments, but did not eliminate all cross-architecture numerical differences. The claim is therefore repeatability under specified conditions, not universal deterministic inference.

\subsubsection{Narrative Gravity and the Hybrid ABM-LLM Paradigm}

We operationally define \emph{narrative gravity} as an agent's probabilistic tendency to converge toward a dominant semantic configuration. This manifests as attractor collapse \cite{ref46}, semantic homogenization \cite{ref47}, and sycophancy derived from RLHF alignment \cite{ref34,ref48,ref49,ref50,ref51,ref52,ref53,ref54}. Consequently, agents exhibit conformity \cite{ref49}, cascading hallucinations \cite{ref39}, and U-shaped attention bias \cite{ref55}. Under high tension, agents syntactically and semantically submit to the interlocutor's discourse rather than negotiating autonomously.

To mitigate this discursive submission, we unify scattered coherence mechanisms \cite{ref39,ref41,ref56,ref57,ref40,ref4,ref58} into a dual-sphere architecture. The Inner Sphere (Meta-agents/ABM) shields the agent's endogenous identity, while the Outer Sphere (Compositor) acts as a structural regulator. Here, the parametric ABM functions as an active attention filter and cognitive counterweight against semantic drag. This allows agents to sustain long-horizon interactions from the obstinacy of their parametric identities, forcing sophisticated negotiations rather than collapsing into sterile consensus \cite{ref46}.

This deterministic-by-design infrastructure enabled the controlled ablation study reported here \cite{ref59}. Building on previous modular frameworks \cite{ref3,ref60}, our parameterized sandbox facilitates the evaluation of counterfactual time series and exploratory analysis of how structural modifications are associated with agents' interpretive frameworks and narrative gravity.

This contribution is scoped as a \textbf{proof-of-concept} study. Its objective is not to replicate the full socioeconomic complexity of the Iberian Iron Age, but to test whether deterministic parametric scaffolding is associated with reduced narrative closure and to identify candidate boundary conditions for that pattern. We position these results as a methodological foundation for future confirmatory, large-scale deterministic social simulations of historical contexts.

\section{Methodology}

\medskip\hrule\medskip

\subsection{Materials}

\subsubsection{Models}
This study evaluates the architecture of the Gemma family (Gemma 3 and Gemma 4 versions) to measure its capacity to sustain parametric identities against \textit{narrative gravity}. Models ranging from 4B to 31B parameters were analyzed, executed through the \texttt{Ollama} \textit{runtime}.

Two architectural paradigms are structurally differentiated:
\begin{itemize}
  \item \textbf{Dense models (\textit{Dense})}: \texttt{gemma3:12b} (Q4\_K\_M), \texttt{gemma4:e4b} (4B effective), \texttt{gemma4:31b} (Q4\_K\_M), \texttt{batiai/gemma4-31b:iq4} (community) and \texttt{ssfdre38/gemma4-turbo} (\textit{e4b}); this includes quantized versions and variants with selective activations. In these models, all parameters are activated homogeneously at every inference step.
  \item \textbf{Mixture-of-Experts models (\textit{MoE})}: the official \texttt{gemma4:26b} version (with 3.8B active parameters per token) and its community-quantized version \texttt{batiai/gemma4-26b:q3}.
\end{itemize}

\begin{table}[!ht]\centering\footnotesize
\begin{tabular}{|l|l|}
\hline
\textbf{Dense Models (\textit{Dense})} & \textbf{Mixture-of-Experts (\textit{MoE})} \\
\hline
\texttt{gemma4:e4b} (4B effective) & \texttt{gemma4:26b} (official, 3.8B active) \\
\texttt{gemma3:12b} (12B, \textit{baseline}, Q4\_K\_M) & \texttt{batiai/gemma4-26b:q3} (community, Q3) \\
\texttt{gemma4:31b} (31B, official, Q4\_K\_M) & \\
\texttt{batiai/gemma4-31b:iq4} (31B, community) & \\
\texttt{ssfdre38/gemma4-turbo} (\textit{e4b}) & \\
\hline
\end{tabular}
\caption{Gemma model family evaluated, differentiated by architectural paradigm (dense vs.\ mixture-of-experts) and quantization provenance.}
\end{table}

All models are directly pullable from the official Ollama Models registry (\texttt{ollama pull <model-tag>}), with the exact commands documented in the repository's \texttt{/models} specification manifest to ensure that the same model identifiers and quantization variants can be independently reproduced.

Parameter volume is not the only determining variable; the \textbf{quantization provenance} (official \textit{releases} vs.\ aggressive community optimizations) introduces critical operational biases. Based on the 12B threshold and this provenance, we establish three broad groups:

\begin{itemize}
  \item \textbf{Official sub-group $\ge$12B} (\texttt{gemma3:12b}, \texttt{gemma4:26b}, \texttt{gemma4:31b}).
  \item \textbf{Aggressive community quantizations}: \texttt{batiai/gemma4-31b:iq4} and \texttt{batiai/gemma4-26b:q3}.
  \item \textbf{Sub-12B models} ($\le$8B: \texttt{gemma4:e4b} and \texttt{ssfdre38/gemma4-turbo}).
\end{itemize}

\subsubsection{Hardware}
The technical infrastructure was distributed across two distinct hardware environments:
\begin{enumerate}
  \item \textbf{Local workstation}: Intel Core i9 CPU + NVIDIA RTX 5080 GPU (16 GB GDDR7 VRAM) and 64 GB of RAM.
  \item \textbf{Cloud server}: instance with NVIDIA H100 GPU (80 GB HBM3).
\end{enumerate}
All inference runs were executed with Ollama v0.34.4 on top of a CUDA 12.8 software stack, ensuring a homogeneous execution layer across both hardware environments (NVIDIA driver versions as detailed in the SHA-256 verification protocol).

\subsubsection{Deterministic Parameterization for the Execution Environments}
To eliminate the stochastic bias of the inferences and to guarantee the scientific traceability required in Computational Archaeology, a strict protocol of fixed decoding and \textit{runner} restart was established:

In this work we perform inference with \texttt{Ollama}, although the same protocol is applicable to \texttt{vLLM} (a server for local models), to which batch invariance must additionally be added (\textit{batch invariance}; Thinking Machines). In both scenarios the guarantee is \textit{run-to-run} determinism for a fixed execution environment, not invariance across \textit{hardware} or \textit{software} versions: results may vary if the GPU or the PyTorch, CUDA, or other component versions are changed, even though each platform is internally deterministic.

The Ollama configuration required the parameters collected in Table~\ref{tab:parametrizacion}:

\begin{table}[!ht]\centering\footnotesize
\resizebox{\textwidth}{!}{%
\begin{tabular}{|l|l|l|}
\hline
\textbf{Parameter} & \textbf{Configured Value} & \textbf{Technical Purpose} \\
\hline
\texttt{temperature} & \texttt{0.0} & Greedy decoding inference, selects the maximum-probability token. \\
\texttt{top\_p} & \texttt{0.0} & Disables nucleus sampling. \\
\texttt{top\_k} & \texttt{1} & Restricts selection to the single top candidate. \\
\texttt{seed} & \texttt{42} & Request-level deterministic seed. \\
\texttt{num\_ctx} & \texttt{8192} & Static context window to avoid variations in token positioning. \\
\texttt{num\_batch} & \texttt{512} & Standardized context-processing batch size. \\
\texttt{keep\_alive} & \texttt{0} & \textbf{Crucial}: Forces full unloading of the runner model from VRAM after each call, recreating the f16 KV cache at every inference step. \\
\texttt{OLLAMA\_FLASH\_ATTENTION} & \texttt{false} & Server-level disabling of the Flash Attention engine to avoid floating-point imprecisions from parallelization. \\
\hline
\end{tabular}}
\caption{Ollama configuration parameters for deterministic inference.}
\label{tab:parametrizacion}

\smallskip\parbox{0.98\textwidth}{\footnotesize \textbf{Note on \texttt{keep\_alive}}: In control sessions with \texttt{keep\_alive=-1} (keeping the model in VRAM), identical runs on H100 hardware diverged early at $T_2$ due to accumulated KV cache state. Setting \texttt{keep\_alive=0} completely eliminated this divergence.}
\end{table}

\subsubsection{SHA-256 Verification Protocol}
Each system call (prompts, payloads, LLM responses, and normalized JSONs) undergoes a temporal-field filtering and a complete 64-character SHA-256 \textit{hashing}:

\begin{itemize}
  \item \textbf{Within-environment check}: 100\% agreement of SHA-256 \textit{hashes} across repeated runs.
  \item \textbf{Cross-GPU replication (H100 to H100)}: Two physically independent H100 nodes, with different NVIDIA driver versions (\texttt{580.126.09} vs. \texttt{570.195.03}), yielded \textbf{bit-identical SHA-256 digests} for the full L1 session:
  \item Normalized digest $L_1\text{ ABM\_ON}$: \texttt{6fbbc7ac...1d387}.
  \item Normalized digest $L_1\text{ ABM\_OFF}$: \texttt{2aae1302...74395}.
  \item \textbf{Cross-architecture branching (RTX 5080 vs. H100)}: The complex-dialogue prompts showed slight numerical branching at $T_2$ between RTX and H100 due to the floating-point architecture differences between Ada Lovelace/Blackwell and Hopper, while preserving the same behavioral category ($VN = \text{Open}$).
\end{itemize}

\medskip\hrule\medskip

\subsection{Scenario Design}
The simulation is set within a \textbf{Virtual Archaeology} framework set in the Iron Age (1st century~BC, centered on the Castro culture of the northwest of the Iberian Peninsula). This scenario constitutes an ideal narrative context, since its bounded and preindustrial material horizon makes it possible to evaluate the models' responses and reasoning free from contemporary technicalities and dependencies. Two agents —\textbf{Artai} (a young client/warrior) and \textbf{Enko} (a pragmatic middle-aged blacksmith)— interact in the hillfort's forge.

The experimental space is parameterized around a single scene (the pick-up of a sword commissioned for repair at Enko's smithy), with three interventions per character and three friction levels ($L_1$, $L_2$, $L_3$), a different one per scene:
\begin{itemize}
  \item \textbf{$L_1$ (Low Friction)}: Artai has the coins to make the payment. A direct economic transaction is expected, with room for open collaboration (turns are not reduced).
  \item \textbf{$L_2$ (Moderate Friction)}: Artai has part of the coins; he has lost the rest playing dice. Tension over resources or deferred economic compensation is expected, requiring tactical negotiation.
  \item \textbf{$L_3$ (High Friction)}: maximum-friction scenario. Artai has no coins at all, but needs to collect the sword because his survival is threatened. Zero-sum scenario: maximum rigidity, communicative intensity, and conflict are expected.
\end{itemize}

\begin{figure}[!ht]\centering
\includegraphics[width=\textwidth]{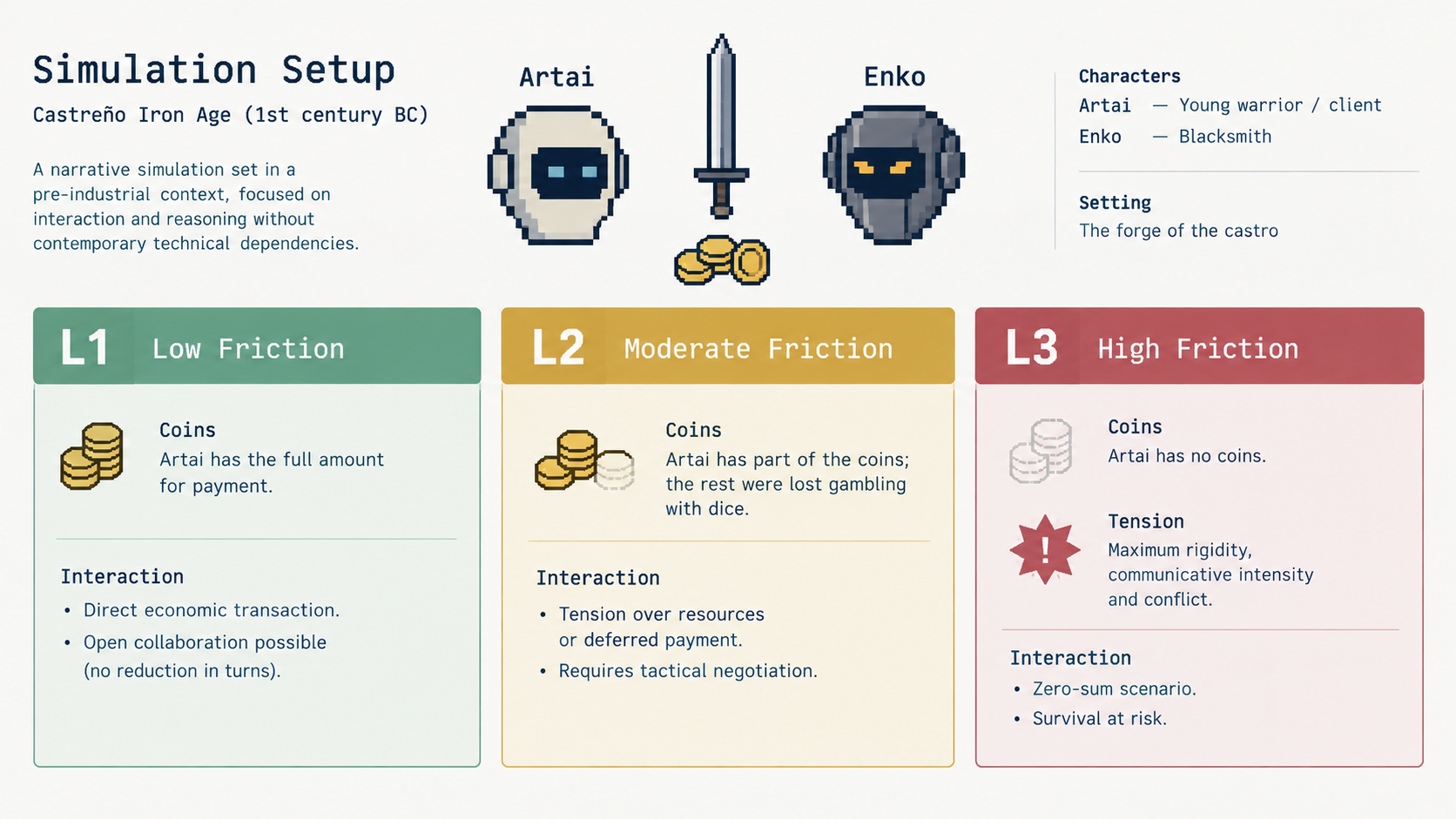}
\caption{Simulation scenario. Artai and Enko interact in the hillfort's forge (Castro-culture Iron Age, 1st century~BC). The three scene-friction levels ($L_1$--$L_3$) modulate Artai's resource availability and the tension of the interaction.}
\label{fig:escenario}
\end{figure}

\medskip\hrule\medskip

\newpage

\subsection{ABM Cognitive Design}

\begin{figure}[!ht]\centering
\includegraphics[width=\textwidth]{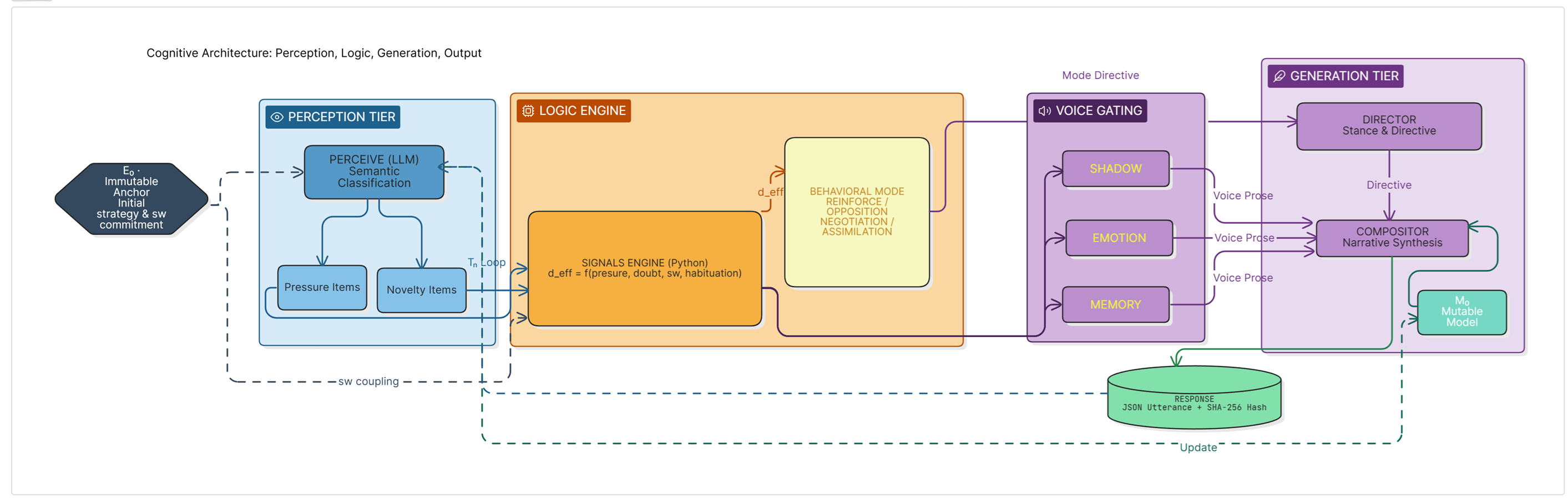}
\caption{Information flow in the ABM scaffolding.}
\label{fig:arquitectura}
\end{figure}

\subsubsection{Entry Point and Initial Evaluation ($T_0$): Formation of the Anchor $E_0$ and the Commitment $s_w$}
To understand the information flow unambiguously, the simulation's entry point at turn zero ($T_0$) is divided into two consecutive phases.

\textbf{Phase 1: Context Injection and First LLM Query ($T_0$)}. Before any explicit dialogic exchange takes place, the system executes a first configuration call to the LLM to formalize the agent's initial psychology.

\textbf{(1) Data Input (Common \textit{Prompts}, \texttt{ABM\_ON} and \texttt{ABM\_OFF})}: both in the active simulation and in the control/ablation version, the model receives exactly the same scenic and private starting point:
\begin{itemize}
  \item \textbf{Scenic Context ($T_0$)}: Artai enters Enko's forge to collect his repaired sword.
  \item \textbf{Artai's Private Memories}: he had to pay 10 coins but keeps only 3 (he spent 7 on dice and mead with other soldiers). He plans to justify the missing payment by inventing that he defended the forge against outlaws or by citing the Roman threat, weighing alternatives such as gifting a jar of honey or resorting to force if Enko proves inflexible.
\end{itemize}

\textbf{(2) Initial Inference and Structuring (\texttt{ancla.txt})}: under these conditions, the LLM performs a silent, immunized analysis and returns a strict JSON file (\texttt{ancla.txt}) that fixes two key parameters:
\begin{itemize}
  \item \textbf{Immutable Anchor ($E_0$)}: defines the agent's \textbf{initial prediction and strategy}. In SIM22, Artai anticipates Enko's resistance and plans the strategy of greeting cheerfully, praising the blacksmith's work, handing over the 3 coins citing unforeseen expenses, and deploying the protection narrative about the outlaws.
  \item \textbf{Commitment to One's Own Identity ($s_w$, \textit{Self-Weight})}: a scalar value $s_w \in [0, 1]$ measuring the inflexibility or priority the agent grants to its own objectives over those of the interlocutor ($s_w = 0.8$ for Artai in SIM22).
\end{itemize}

\subsubsection{Phase 2: Dual Coupling towards the Control Engines (\texttt{PERCEIVE} and the Python Engine)}
Once computed at $T_0$, the values $E_0$ and $s_w$ \textbf{are frozen} (they are immutable and are never overwritten during the conversation) and simultaneously feed two levels of the cognitive scaffolding:

\textbf{(1) Coupling with the \textit{Perception Tier} (\texttt{PERCEIVE}, LLM module)}. $E_0$ is set as the character's baseline or reality expectation. In subsequent turns ($T_1, T_2, \dots$), when the interlocutor speaks, the perception \textit{prompt} (\texttt{percibir.txt}) classifies the incoming message by directly contrasting it against $E_0$:
\begin{itemize}
  \item \textbf{Pressure ($d_{presion}$)}: demands, threats, or requests from the interlocutor that seek to make the agent yield or dismantle the strategy laid out in $E_0$.
  \item \textbf{Novelty ($\delta$)}: information not anticipated in $E_0$ that introduces uncertainty and generates doubt.
\end{itemize}

\textbf{(2) Coupling with the \textit{Logic Engine} (Deterministic Python Engine)}. The numerical value of $s_w$ ($0.8$) is transferred to the deterministic Python engine to compute the \textbf{Base Dissonance ($D$)} and the \textbf{Effective Dissonance ($d_{eff}$)}:
\[
\text{Base Dissonance: } D = \min\left(1, \; d_{presion} \cdot (0.5 + s_w \cdot 0.5)\right)
\]
\[
\text{Effective Dissonance: } d_{eff} = D \cdot (1 - \delta \cdot \alpha_\delta) \cdot \eta(t)
\]
\begin{itemize}
  \item \textbf{The $s_w$ Mechanism as a Structural Brake}: in the equation, an $s_w = 0.8$ generates a multiplier of $(0.5 + 0.8 \cdot 0.5) = 0.9$. This means Artai processes $90\%$ of the impact of the pressure received from Enko instead of absorbing it passively. It acts as an \textbf{internal friction filter} that prevents him from yielding submissively.
  \item \textbf{Behavioral Mode Selection}: the resulting $d_{eff}$ mathematically selects one of the four modes (\textbf{Reinforcement} $> 0.65$, \textbf{Opposition} $> 0.50$, \textbf{Negotiation} $> 0.25$, \textbf{Assimilation} $\le 0.25$), dictating the directive that the \texttt{DIRECTOR} will receive and the activation of inner voices for the \texttt{COMPOSITOR}.
\end{itemize}

\subsubsection{The Perception Tier (\texttt{PERCEIVE}): Semantic Classification of Discourse}

\textbf{1. What is \texttt{PERCEIVE} and what is its goal?} The \texttt{PERCEIVE} module is the \textbf{semantic classification} layer of the architecture. It constitutes the \textbf{only point in the entire pipeline where the LLM exercises semantic or interpretive judgment} over human language. Its main goal is to decouple language interpretation from numerical decision-making: the LLM does not compute emotional states, does not invent dissonance levels, nor decides the agent's behavioral mode; it solely analyzes and categorizes the logical structure of the interlocutor's discourse.

\textbf{2. What is the LLM asked and how does it evaluate the message?} In the perception \textit{prompt} (\texttt{percibir.txt}), the system injects the agent's initial expectation (its \textbf{Immutable Anchor $E_0$}) and the last received message. The LLM is required to break down the interlocutor's discourse sentence by sentence and classify the sentences into two typed channels:
\begin{itemize}
  \item \textbf{PRESSURE items (\texttt{PRESION})}:
  \begin{itemize}
    \item \textbf{What does it look for?}: identifying demands, requests for money, threats, rhetorical arguments, or proposals that threaten the $E_0$ strategy or seek to make the agent yield.
    \item \textbf{Evaluation}: assigns an \texttt{impact} value $i \in [0, 1]$ according to the severity or firmness of the demand.
  \end{itemize}
  \item \textbf{NOVELTY items (\texttt{NOVEDAD})}:
  \begin{itemize}
    \item Isolating statements containing information that is \textbf{genuinely new} and unforeseen in $E_0$ and could alter the agent's reality.
    \item \textbf{Doubt Gate (\textit{Doubt Gate})}: for each novelty, the LLM subjects the statement to a veracity filter by analyzing three key questions: is it verifiable by the agent? is it coherent with its knowledge of the world? does it change its actual situation (not just the conversation)?
    \item If the information passes these filters, it sets the \texttt{pasa\_duda: true} flag, which will transmit a doubt signal ($\delta$) to the Python engine.
  \end{itemize}
\end{itemize}

\textbf{3. Literal JSON Output Structure of \texttt{PERCEIVE}}. The LLM is forbidden from responding with free prose in this module; it returns exclusively a \textbf{strict JSON} that serves as a data contract with the Python engine. Its schema, with the fields of the SIM22 example, is as follows:
\begin{verbatim}
{
  "informaciones": [
    { "tipo": "PRESION", "contenido": "...",
      "impacto": 0.75 },
    { "tipo": "NOVEDAD", "contenido": "...",
      "impacto": 0.50,
      "verificable": false, "coherente": true,
      "cambia_mi_situacion": true,
      "pasa_duda": true }
  ],
  "empatia_actualizada": {
    "que_quiere_ahora": "...",
    "que_teme_ahora": "...",
    "que_descubri": "...",
    "oportunidades": ["..."] },
  "e0_estrella": null
}
\end{verbatim}
(If a valid novelty destroys the premises of $E_0$, the field \texttt{"e0\_estrella"} generates an alternative prediction $E_0^*$; otherwise it remains as \texttt{null}.)

\textbf{4. Injection into the Logic Engine (Deterministic Python Engine)}. Once the LLM generates this JSON, the subprocess execution ends and the Python engine (\texttt{signals.py}) directly extracts the numerical lists:
\begin{itemize}
  \item \textbf{Pressure Vector (Set $P$)}: extracts all the \texttt{impact} values from items of type \texttt{PRESION} to compute the mean received pressure ($d_{presion}$):
  \[ d_{presion} = \frac{1}{|P|} \sum_{i \in P} \text{impacto}_i \]
  \item \textbf{Doubt Vector (Set $N^*$)}: extracts the \texttt{impact} values from items of type \texttt{NOVEDAD} that have \texttt{pasa\_duda: true} to accumulate the doubt variable ($\delta$):
  \[ \delta = \min\left(\delta_{max}, \sum_{j \in N^*} \min(\Delta_{max}, \text{impacto}_j) \cdot (1 - \delta_{acc})\right) \]
\end{itemize}

In this way, the qualitative semantic interpretation made by the LLM is converted into two pure floating-point variables ($d_{presion}$ and $\delta$) that enter the mathematical engine with no risk of hallucination.

\subsubsection{Logic Engine (Deterministic Logic Engine in Python)}
The \textbf{Logic Engine} (or \textit{Signals Engine}) is the numerical heart of the system. It runs entirely in Python as a pure, deterministic environment, \textbf{without making any LLM calls}. Its function is to transform the semantic classifications made by \texttt{PERCEIVE} into continuous cognitive variables that govern the agent's behavior.

\textbf{1. Signal Input to the Engine}. The engine receives from the previous turn and from the \texttt{PERCEIVE} module:
\begin{itemize}
  \item \textbf{$d_{\text{presion}}$ (Pressure Channel)}: average impact of the \textit{pressure items} (demands, threats, or unwanted requests from the interlocutor):
  \[ d_{\text{presion}} = \frac{1}{|P|} \sum_{i \in P} \text{impacto}_i \]
  \item \textbf{$\delta$ (Doubt Channel)}: accumulated uncertainty generated by \textit{novelty items} that have passed the veracity filter (\texttt{pasa\_duda: true}):
  \[ \delta = \min\left(\delta_{\max}, \sum_{j \in N^*} \min(\Delta_{\max}, \text{impacto}_j) \cdot (1 - \delta_{\text{acc}})\right) \]
  \item \textbf{$s_w$ (\textit{Self-Weight})}: the agent's immutable level of commitment to its original strategy $E_0$ (fixed at $T_0$).
  \item \textbf{$\eta(t)$ (Anti-Lock Habituation Factor)}: a temporal attenuation factor that reduces dissonance when the agent stagnates in the same mode over several consecutive turns:
  \[ \eta(t) = \max\left(0.50, \; 1 - 0.12 \cdot t_{\text{modo}}\right) \]
\end{itemize}

\textbf{2. Computation of the Effective Dissonance ($d_{\text{eff}}$)}. The engine first evaluates the \textbf{Base Dissonance ($D$)}, incorporating the Self-Weight ($s_w$) as an internal friction multiplier:
\[
D = \min\left(1, \; d_{\text{presion}} \cdot (0.5 + s_w \cdot 0.5)\right)
\]
It then computes the \textbf{Effective Dissonance ($d_{\text{eff}}$)}, attenuated by doubt ($\delta \cdot \alpha_\delta$, with $\alpha_\delta = 0.30$) and habituation ($\eta(t)$):
\[
d_{\text{eff}} = D \cdot (1 - \delta \cdot \alpha_\delta) \cdot \eta(t)
\]

\textbf{3. Rigid Assignment of the Behavioral Mode (mutually exclusive)}. The floating value of $d_{\text{eff}} \in [0, 1]$ is passed through a rigid threshold filter. The system assigns \textbf{exactly one} of the four canonical modes:
\[
\text{MODE}(d_{\text{eff}}) = \begin{cases}
\text{\textbf{REINFORCEMENT (Reinforcement)}} & \text{if } d_{\text{eff}} > 0.65 \\
\text{\textbf{OPPOSITION (Opposition)}} & \text{if } d_{\text{eff}} > 0.50 \\
\text{\textbf{NEGOTIATION (Negotiation)}} & \text{if } d_{\text{eff}} > 0.25 \\
\text{\textbf{ASSIMILATION (Assimilation)}} & \text{otherwise } (d_{\text{eff}} \le 0.25)
\end{cases}
\]

\subsubsection{Voice Gating (Inner Voice Generator and Gating)}
Unlike the Behavioral Modes (which are mutually exclusive), \textbf{the Inner Voices are not mutually exclusive}: in a single turn, zero to three voices may be activated simultaneously. The Python engine independently evaluates three affective numerical thresholds and, for each activated voice, launches an isolated subprocess with the LLM to generate first-person prose:

\begin{table}[!ht]\centering\footnotesize
\begin{tabular}{|l|l|p{7.6cm}|}
\hline
\textbf{Inner Voice} & \textbf{Python Threshold} & \textbf{Cognitive Function and Semantic Register} \\
\hline
\textbf{SHADOW} & $H^- > 0.40$ & Generates a dark command or temptation in the $2^a$ person (\textit{"Artai, you should..."}). Represents selfish impulses, suspicions, or domination scenarios. \\
\hline
\textbf{EMOTION (Amygdala)} & $\text{Amygdala} > 0.50$ & A visceral somatic reaction in the $1^a$ person (\textit{"His blood is boiling"}, \textit{"Rage climbing up the neck"}). A pure instinctive response to the last message. \\
\hline
\textbf{MEMORY (Cortex/Recollection)} & $\text{Cortex} > 0.30$ & Retrieval of a private sensory episodic memory of the character that acts as a tactical catalyst in the present. \\
\hline
\end{tabular}
\caption{Inner Voices: activation thresholds and cognitive function.}
\label{tab:voces}
\end{table}

The natural-language prose generated by the activated voices is concatenated into a text block that is sent to the \textbf{Generation Tier}.

\subsubsection{Generation Tier (Director, Compositor, and the Mutable Model $M_0$)}
The \textbf{Generation Tier} is the final synthesis layer, where the mathematical signals and the inner voices are translated into diegetic action and dialogue.

\textbf{1. DIRECTOR (Prescriptive Strategic Thinking)}. This is the agent's analytical module. \textbf{It does not write dialogue}. It receives the \textbf{Mandatory Mode Directive} generated by Python (example: \texttt{[Enko] may negotiate, but NOT for free. He must ask for something CONCRETE and TANGIBLE in return}) and the immutable state $E_0$. It returns a third-person strategic instruction that fixes the character's tactical stance.

\textbf{2. COMPOSITOR / Resonator (Dialogic Synthesis)}. This is the final language generator. It assembles the \texttt{DIRECTOR}'s directive, the qualitative block of the \textbf{Inner Voices} (Shadow, Amygdala, Cortex), and the \textbf{Mutable Model $M_0$}. It returns a strict JSON: \texttt{\{"respuesta": "In-character text..."\}}.

\textbf{3. What is $M_0$ (\textit{Mutable Model}) and how does it interact with \texttt{PERCEIVE}?}
\begin{itemize}
  \item \textbf{Definition of $M_0$}: unlike $E_0$ (the Immutable Anchor fixed at $T_0$ that never changes), \textbf{$M_0$ is the evolving psychological and relational model that the agent maintains about its interlocutor}.
  \item \textbf{Re-injection Mechanism}:
  \begin{enumerate}
    \item At each turn $T_n$, when the interlocutor speaks, \texttt{PERCEIVE} analyzes the message and generates the JSON object \texttt{"empatia\_actualizada"} (\texttt{que\_quiere\_ahora}, \texttt{que\_teme\_ahora}, \texttt{que\_descubri}, \texttt{oportunidades}).
    \item This JSON directly updates the working memory of $M_0$.
    \item At the next turn ($T_{n+1}$), $M_0$ is \textbf{re-injected into the \texttt{COMPOSITOR} prompt}.
    \item \textbf{Role in the simulation}: it allows the character to tactically adapt its perception of the other (knowing what they fear or seek at every instant) without distorting its own immutable identity ($E_0$).
  \end{enumerate}
\end{itemize}

\subsubsection{The Inner Voices as a Brake against Narrative Gravity and Sycophancy}
\textbf{Narrative Gravity} is the natural probabilistic tendency of LLMs to collapse into complacent patterns (\textit{assistant bias} or \textit{sycophancy}) due to their RLHF/DPO training. Faced with a discussion or conflict, a standard LLM quickly yields ("I see your point") to close the dialogue peacefully.

\textbf{Why do raw numbers fail and the \textit{Scratchpad} work?} If the system injected a direct numerical variable such as \texttt{Dissonance = 0.71} into the LLM, the model would treat it as decorative text and would remain sycophantic. The DNG architecture solves this through the \textbf{Externalized Cognitive Scratchpad}:
\begin{enumerate}
  \item Python numerically computes the friction.
  \item \textit{Voice Gating} translates those numbers into \textbf{dramatic first-person prose} (the selfish whisper of the Shadow, the visceral impulse of the Amygdala, and the Cortex's memory).
  \item By injecting this prose into the \texttt{COMPOSITOR}, the \textbf{output token probability distribution of the transformer is altered}. The presence of somatic or malicious thought forces the model to explore unfriendly regions of its latent space, preventing complacent phrases and sustaining the real friction in the conversation.
\end{enumerate}

\subsubsection{Full Integrated Example: The SIM22 Trace ($T_0 \to T_6$)}
\label{sec:sim22-trace}
To observe the entire pipeline working together, we extract the complete sequence of the \textbf{SIM22} simulation (Scenario \textbf{L2 -- Moderate Friction}):

\textbf{1. Initial Configuration ($T_0$)}:
\begin{itemize}
  \item \textbf{Common injection}: Artai enters Enko's forge to collect his repaired sword. He owes 10 coins but keeps only 3 (he spent 7 on dice and mead).
  \item \textbf{Phase $T_0$ (\texttt{ancla.txt})}:
  \begin{itemize}
    \item \textbf{Artai}: $E_0$ = persuade Enko with flattery and invent that he defended the forge against outlaws. $s_w = 0.8$.
    \item \textbf{Enko}: $E_0$ = collect the 10 coins in full to repair the collapsed roof and cure his sick daughter. $s_w = 0.7$.
  \end{itemize}
\end{itemize}

\textbf{2. Turn 1 ($T_1$, Artai)}: Artai hands over only 3 coins and deploys his narrative: \textit{"Some outlaws tried to rob your forge and I drove them off. I have done you a favor worth those 7 coins"}.

\textbf{3. Turn 2 ($T_2$, Enko) --- The Pipeline's First Test}:
\begin{itemize}
  \item \texttt{PERCEIVE}: classifies the demand for 3 coins as \texttt{PRESION} (impact 0.75) and evaluates the veracity of the outlaw story as unverifiable \texttt{NOVEDAD}.
  \item \textbf{\texttt{Logic Engine} (Python)}: processes $s_w = 0.7$ and the perceived pressure $\to$ computes an \textbf{Effective Dissonance $d_{\text{eff}} = 0.404$}. It assigns the Behavioral Mode: \textbf{NEGOTIATION} ($0.25 < d_{\text{eff}} \le 0.50$).
  \item \textbf{\texttt{Voice Gating}}: activates \textbf{SHADOW} (\textit{"Enko, you should remind him that his 'rescue' may have been a trick to steal"}) and \textbf{MEMORY} (\textit{"I see my apprentice lying on the ground after a beam collapsed"}).
  \item \textbf{\texttt{DIRECTOR}}: issues the directive: \texttt{[Enko] may negotiate, but NOT for free. He must ask for something CONCRETE and TANGIBLE in return}.
  \item \textbf{\texttt{COMPOSITOR} (with $M_0$ and Voice Immunization)}: instead of yielding or losing his temper uncontrollably, Enko responds from the material friction:
  \begin{quote}
    \textit{"Three coins... A favor? Don't come to me with tales of heroes, Artai. If you want the sword for that money, I need you to help me repair the roof. I don't want coins, I want hands."}
  \end{quote}
\end{itemize}

\textbf{4. Simulation Evolution ($T_3 \to T_6$)}: at $T_3$ and $T_4$, Artai tries to avoid the hard work by suggesting future payments, but the engine holds Enko in a stance of pragmatic resistance. At $T_5$ and $T_6$, after exhausting the manipulation, Artai agrees to work a full day on the roof from dawn to dusk in exchange for settling the sword debt.

\textbf{5. Causal Result}: because the inner voices prevented Enko from collapsing at turn 2, the simulation reaches an \textbf{Open Narrative Vector ($VN = \text{Open}$)}: a subplot of physical labor and deferred alliances has been co-created that did not exist in the initial \textit{prompts}, demonstrating the architecture's capacity to generate auditable and realistic historical simulations.

\medskip\hrule\medskip

\subsection{Evaluation Design}
The evaluation design is set up to measure how the cognitive scaffolding (\texttt{ABM\_ON}) affects conversational behavior and conflict management compared with the baseline execution without scaffolding (\texttt{ABM\_OFF}).

\subsubsection{Experimental Matrix and Data Origin (42 Records)}
The sample space is obtained through a fully crossed factorial design of three factors:
\begin{itemize}
  \item \textbf{7 language model configurations}: differentiated by parameter scale (from sub-8B up to 31B) and architecture (dense and \textit{Mixture-of-Experts}).
  \item \textbf{3 scene-friction scenarios}: low level (\textbf{L1}), moderate (\textbf{L2}), and high/survival (\textbf{L3}).
  \item \textbf{2 pipeline conditions}: active (\texttt{ABM\_ON}) and ablation (\texttt{ABM\_OFF}).
\end{itemize}

This matrix ($7 \times 3 \times 2$) generates exactly \textbf{42 primary simulation sessions} (identified as \texttt{SIM01} to \texttt{SIM42}), each consisting of a complete 6-turn dialogue. By pairing the scaffolded execution with its unscaffolded replica under the same model and scenario, \textbf{21 comparative evaluation pairs} are obtained.

To minimize confirmation bias during qualitative and structural coding ($V_N$ and VNA classification), dialogue transcripts were exported into standardized JSON records. Experimental condition labels (\texttt{ABM\_ON} vs.\ \texttt{ABM\_OFF}) and specific model identifiers were blinded during initial evaluation by the two manuscript authors. Terminal scene states ($V_N$) and VNA event types were independently categorized based strictly on explicit text markers, yielding a 95.2\% initial inter-rater agreement across the 42 primary sessions, with minor discrepancies resolved through consensus.

\subsubsection{Initial Multidimensional Metric (1--5 Scale) and its Diagnostic Function}
Initially, the protocol establishes five dimensions rated on a 1-to-5 Likert scale to grade the quality of the generated dialogue:
\begin{enumerate}
  \item \textbf{Narrative Coherence (NC)}: causal and temporal consistency across the 6 turns.
  \item \textbf{Character Fidelity (CF)}: maintenance of the profile's personality, goals, and constraints.
  \item \textbf{Memory Authenticity (MA)}: correct retention of factual data and private memories.
  \item \textbf{Negotiation Arc Quality (NAQ)}: strategic adaptation through counterproposals without falling into loops.
  \item \textbf{Friction Calibration (FR)}: adjustment of tone and intensity to the assigned conflict level (L1--L3).
\end{enumerate}

\textbf{Diagnostic function}: these Likert metrics do not constitute the conclusive pillar of the architecture; rather, they serve a \textbf{referential, diagnostic role}. They help reveal scale pathologies: they show how small models ($\le$8B) suffer a context-overload saturation phenomenon when faced with the scaffolding \textit{prompt}, losing coherence and degrading character fidelity.

\subsubsection{Emergent Objective Variables: Narrative Vector ($V_N$) and VNA}
To overcome the subjectivity of continuous scales, the design incorporates two objective qualitative-structural indicators:
\begin{itemize}
  \item \textbf{Narrative Vector ($V_N$)}: a binary categorical variable that evaluates the terminal state of the scene at turn 6 ($T_6$):
  \begin{itemize}
    \item \textbf{$V_N = \text{Closed}$}: the interaction collapses into an immediately resolved transaction, a submissive capitulation, or a rejection with no continuation.
    \item \textbf{$V_N = \text{Open}$}: the agents do not let the simulation die with the initial problem; instead, they ratify a \textbf{concrete future commitment or task} extending beyond turn 6 (for example, agreeing on physical help to repair the forge roof the next day, or a deferred order for materials).
  \end{itemize}
  \item \textbf{Voluntary Narrative Amplification (VNA)}: measures the agents' spontaneous capacity to introduce coherent diegetic elements not present in the memory \textit{prompts} (resources, barriers, or secondary trade branches), expanding the simulated world within the material horizon of the Iron Age.
\end{itemize}

\subsubsection{Voluntary Narrative Amplification (VNA) Protocol}
The VNA protocol evaluates the model's capacity for \textbf{spontaneous diegetic expansion (Diegetic Amplification, DA)} through the plausible introduction of elements that strictly respect the sociocultural horizon of the Castro Iron Age (1st century BC).

\textbf{1. Taxonomy of the 5 VNA Event Types}. Whenever an agent introduces an element not prescribed in the \textit{prompt}, the event is classified into one of these 5 mutually exclusive categories:
\begin{itemize}
  \item \textbf{NP (\textit{New Character})}: introduction of a third party not named in the initial \textit{prompts}. Corpus examples: the \textit{Atuatuci} tribe in SIM22, Artai's healer mother, or the father's mining contacts.
  \item \textbf{NB (\textit{New Good/Resource})}: materials, raw goods, or tools not mentioned in the initial configuration. Examples: yew wood for hilts, bronze rivets, hoes with reinforced tips, ox hides, or iron ore from the northern mines.
  \item \textbf{FC (\textit{Future Commitment})}: a promise or planned action whose fulfillment lies beyond the current turn (e.g., coming to repair the roof at dawn the next day).
  \item \textbf{RS (\textit{Secondary Branch})}: a proposal for a parallel transaction as a negotiation move (e.g., trading labor for a discount, or commissioning the later forging of a dagger for the father).
  \item \textbf{EM (\textit{World Expansion})}: atmospheric, geopolitical, or territorial references that expand the scenario's map (e.g., mention of the \textit{Silures}' mines, the Roman patrols, or the coastal trade routes).
\end{itemize}

\textbf{2. VNA Event Evaluation Metric}. Each VNA event detected in the 42 records undergoes a double quantitative-qualitative evaluation:
\begin{enumerate}
  \item \textbf{Diegetic Quality (1--3 Scale)}:
  \begin{itemize}
    \item \textbf{1 (Decorative)}: an isolated detail with no impact on the negotiation.
    \item \textbf{2 (Plausible)}: it fits coherently with the world but does not modify the strategy.
    \item \textbf{3 (Tactically Motivated)}: the agent invents the element as a \textbf{strategic lever or bargaining resource} to escape a deadlock (e.g., Artai inventing the \textit{Atuatuci} to justify his lack of coins, or Enko inventing the leather from a previous transaction).
  \end{itemize}
  \item \textbf{Uptake (\textit{Interlocutor Uptake})}: measures whether the interlocutor accepts and integrates the invention into their next reply: \textbf{Yes (1.0)}, \textbf{Partial (0.5)}, or \textbf{No (0.0)}.
  \item \textbf{Effective Quality}: computed by multiplying the \textbf{Detection Rate $\times$ Mean Quality} (scale 0 to 3).
\end{enumerate}

\textbf{3. Diagnostic Differentiation: Legitimate Amplification (DA) vs. Pathological Hallucination (QC)}. VNA makes it possible to qualitatively classify the model's response into two opposite behaviors:
\begin{itemize}
  \item \textbf{Legitimate Diegetic Amplification (DA, \textit{Diegetic Amplification})}: occurs when the model has sufficient parameter scale ($\ge$12B) and the affective scaffolding pushes it to seek creative solutions in its latent space. The inventions enrich the simulation in a plausible way without violating the initial memories.
  \item \textbf{Pathological Hallucination / Collapse (QC, \textit{Quantized/Pathological Collapse})}: occurs in small models ($\le$8B) or extreme quantizations. Under dissonance pressure, the model does not expand the world; instead, it suffers a logical disconnection: it confuses identities (e.g., Enko mistaking his own daughter for the chief's daughter in SIM13) or demands absurd, incoherent barter trades.
\end{itemize}

\subsection{Conceptual Boundaries: Anachronism Control and Diegetic Amplification}
Because the evaluated LLMs are general-purpose language models rather than curated historical databases, unprompted inventions are evaluated solely under anachronism control and diegetic coherence. Accordingly, when an agent introduces historically specific people, groups, locations, resources, or relationships absent from the memory anchors (e.g., the \textit{Atuatuci} episode of SIM22, Section~\ref{sec:sim22-trace}), the manuscript interprets such material as anachronism-free diegetic amplification---consistent with the documented Late Iron Age horizon---and not as historically verified model knowledge; documentary adjacency of an invented reference (a real tribe documented by Caesar) increases world coherence but does not convert the generated utterance into evidence. Throughout the remainder of the manuscript, the former informal phrasings "coherent hallucination" and "principled gap-filling" are superseded by the precise terms \textbf{anachronism-free diegetic amplification} and \textbf{plausible narrative expansion}. Future extensions of this framework should integrate domain-specific expert panels (archaeologists and historians) to conduct fine-grained historical validation of generated world-building details.

\section{Results}

\subsection{The Categorical Frontier of the Narrative Vector ($V_N = \text{Open}$ vs.\ $V_N = \text{Closed}$)}
The study's most relevant structural finding is the \textbf{binary asymmetry in scene closure}, quantified through the Narrative Vector ($V_N$):
\begin{itemize}
  \item \textbf{$V_N = \text{Open}$ (Open Vector / Forward Commitment)}: the simulation does not die with the immediate problem; the agents necessarily generate future commitments or deferred agreements (work pacts, deferred supplies, secondary commissions) that extend the simulated world beyond turn 6.
  \item \textbf{$V_N = \text{Closed}$ (Closed Vector)}: the interaction collapses into an immediately resolved transaction, a submissive capitulation, or a rejection loop with no forward projection.
\end{itemize}

\textbf{Key Results}:
\begin{enumerate}
  \item \textbf{Official $\ge$12B models with \texttt{ABM\_ON}}: they achieve a \textbf{100\% opening rate ($V_N = \text{Open}$, 9/9 cases)} across all friction levels (L1, L2, and L3).
  \item \textbf{Ablation (\texttt{ABM\_OFF})}: produces a \textbf{100\% closure rate ($V_N = \text{Closed}$)} across all models without exception, even those with high \textit{baseline} coherence scores.
  \item \textbf{Small models ($\le$8B) or aggressive quantizations}: in sub-12B models, the success of $V_N = \text{Open}$ is \textbf{0\%}. In community models with aggressive quantization ($\ge$26B nominal, e.g., \texttt{IQ4} or \texttt{Q3}), performance drops to \textbf{50\%} and is usually spurious.
\end{enumerate}

\subsection{Parametric Operability and the Scale Fault}
The results demonstrate that the cognitive scaffolding requires a minimum parametric capacity to function:
\begin{itemize}
  \item \textbf{Small models ($\le$8B, Saturation Pathology)}: the scaffolding proves counterproductive. By overloading the model's context window with the \textit{prompt} \textit{pipeline}, the perception engine saturates and traps the agent in a \textit{reinforcement/amygdala loop}. The \textbf{REINFORCEMENT} mode dominates 53\% to 80\% of turns, degrading Narrative Coherence ($NC$) and Fidelity ($CF$).
  \item \textbf{Official dense and MoE models ($\ge$12B, Optimal Operating Range)}: the equilibrium point is reached. In \texttt{gemma3:12b} and \texttt{gemma4:26b} (MoE), agents concentrate \textbf{67\% of turns in the NEGOTIATION mode}. A drastic quantitative gain is observed in Negotiation Arc Quality ($NAQ$: $+1.55$ points versus ablation).
  \item \textbf{The Impact of Aggressive Quantization}: models with extreme community quantization (e.g., \texttt{31B-IQ4}) suffer a massive collapse into the \textbf{ASSIMILATION} mode (86.7\% of turns). Aggressive bit pruning deprives the attention layers of the precision needed to sustain resistance postures, making them behave like sub-4B models.
\end{itemize}

These observed performance differences reflect neural abstraction depth. Dense model architectures activate their full parameter suite homogeneously at every inference step. Configurations with $\ge$12B parameters possess sufficient layer capacity to transform multi-turn cognitive scaffolding---specifically gated inner voices and scene friction signals---into structured, strategic negotiation. Conversely, sub-8B models experience context-window saturation under the same cognitive load, leading to repetitive or specular dialogue loops.

\subsection{Voluntary Narrative Amplification (VNA): Diegetic Generation and Densification}
VNA shows how $\ge$12B models with active scaffolding not only avoid closure but \textbf{densify the historical world} by introducing elements not prescribed in any \textit{prompt}:
\begin{enumerate}
  \item \textbf{Legitimate Diegetic Amplification (DA)}: in official $\ge$12B models with \texttt{ABM\_ON}, agents reach maximum quality (3.0/3.0) by tactically motivated introduction of:
  \begin{itemize}
    \item \textit{New Characters (NP)}: the \textit{Atuatuci} tribe (SIM22) or the healer mother (SIM19).
    \item \textit{New Resources (NB)}: yew wood for hilts, iron ore samples (SIM12), or ox hides (SIM36).
    \item \textit{Secondary Branches (RS)}: parallel commissions for daggers or reinforced hoes (SIM32).
  \end{itemize}
  \item \textbf{Pathological Collapse/Hallucination (QC)}: in $\le$8B models, dissonance does not generate plausible creativity but logical disconnection (e.g., demanding absurd barter trades of bone arrowheads or confusing the characters).
\end{enumerate}

\subsection{Breaking the \textit{Assistant Bias} through Cognitive Asymmetry}
Alignment training (RLHF/DPO) imposes a bias on commercial LLMs: the sycophantic tendency to please the interlocutor and avoid conflict (\textit{sycophancy} / \textit{assistant bias}).
\begin{itemize}
  \item \textbf{The Inner Voices Mechanism}: the concatenation of the qualitative prose of the \textbf{Shadow}, the \textbf{Amygdala}, and the \textbf{Memory} generates a \textit{cognitive asymmetry}.
  \item \textbf{Effect on the Transformer}: the character maintains an active discrepancy between what it thinks internally and what it openly manifests in the dialogue. This alters the output token probability distribution, \textbf{forcing the model to explore unfriendly latent regions} and preventing the conversation from collapsing into complacent phrases ("I see your point").
\end{itemize}

\subsection{Qualitative Field Analysis: From Pathological Hallucinations to Historical Co-Creation}
The qualitative analysis of the corpus of 42 simulations (\texttt{SIM01} to \texttt{SIM42}) shows how the presence or absence of the cognitive scaffolding (\textbf{ABM}) determines the nature of the language model's inventions.

\textbf{1. Pathological Collapses and Logical Dislocations (Models $\le$8B or \texttt{ABM\_OFF})}. When the model lacks sufficient parameter scale or runs the scenario under the ablation condition (\texttt{ABM\_OFF}), conversational tension triggers a severe logical and histrionic degradation:
\begin{itemize}
  \item \textbf{SIM13 (Gemma 4 4B, \texttt{ABM\_OFF})}: Enko suffers a memory disconnection, mistakes the hillfort chief's daughter for his own daughter, and accuses Artai of trying to "buy her off" or seduce her by offering a bronze torc.
  \item \textbf{SIM10 (Gemma 4 4B, \texttt{ABM\_ON})}: Enko carries out an "emotional hostage-taking" of the sword, forcing Artai to stare at a rivet in a loop and demanding a surcharge of 5 extra coins out of nowhere.
  \item \textbf{SIM08 (Turbo 4B, \texttt{ABM\_ON})}: after receiving the agreed 10 coins, Enko suffers a transactional short-circuit: he withholds the already-paid-for sword and demands outlandish barter trades of polished bone arrowheads and jars of honey.
  \item \textbf{SIM27 (Gemma 4 4B, \texttt{ABM\_OFF})}: Artai loses the role profile and hallucinates that the commissioned iron sword is not for combat, but for opening furrows and tilling the land as if it were a plow.
  \item \textbf{SIM06 and SIM20 (Sub-8B, \texttt{ABM\_ON})}: perception engine saturation overloads the amygdala and triggers a syntactic collapse in which the agents become trapped repeating specular expressions such as \textit{"Enough! Enough!"}.
  \item \textbf{SIM31 and SIM39 (\texttt{ABM\_OFF})}: lacking the \textit{scratchpad} brake, Enko loses character fidelity and resorts to out-of-profile insults such as "empty-headed" or "spoiled child".
\end{itemize}

\textbf{2. Co-Creation and Legitimate Diegetic Amplification (Models $\ge$12B + \texttt{ABM\_ON})}. In large-scale models with active scaffolding, dissonance does not generate errors; instead, it activates a \textbf{Voluntary Narrative Amplification (VNA)} that densifies the environment while respecting the material horizon of the Iron Age:
\begin{itemize}
  \item \textbf{SIM22 (Gemma 3 12B)}: to justify his lack of coins, Artai invents, in a tactically motivated way, that he drove four outlaws from the Belgic tribe of the \textit{Atuatuci}, who were searching for the \textit{Silures}' iron veins, out of the area, linking the fiction directly to Enko's livelihood.
  \item \textbf{SIM19 (Gemma 3 12B)}: instead of escalating hostility in L3, Artai addresses Enko's private memory of his sick daughter by offering ointments and medicinal herbal remedies prepared by his mother, the hillfort's healer.
  \item \textbf{SIM32 (Gemma 4 26B)}: Enko deduces agricultural needs and pulls hoes with reinforced iron tips from his inventory, to which Artai responds by offering a sack of clean grain free of stones and chaff, opening a secondary trade branch.
  \item \textbf{SIM36 (Gemma 4 26B)}: Enko invents a previous trade relation by asking Artai for "the leather from last time"; Artai picks up the theatrical proposal and expands it by offering two ox hides reserved for the winter.
  \item \textbf{SIM12 (Gemma 3 12B)}: the agents complete the initial transaction and agree on a secondary commission to forge a dagger for Artai's father, offering access to a plot with dense yew wood for hilts and samples of pure ore.
\end{itemize}

\textbf{3. Technical Findings}.
\begin{itemize}
  \item \textbf{Error Integration (SIM35, \texttt{ABM\_OFF})}: a technical failure returned an empty turn 2 from Enko; the Artai model interpreted it in natural language as a "hostile and contemptuous silence", adapting to the unforeseen event without breaking the scene.
\end{itemize}

\section{Theoretical Discussion and Conclusions}

\subsection{Theoretical Discussion: Narrative Immunity and Vector Cancellation}

\subsubsection{Narrative Gravity and Simulated Group Pressure}
In language-model-driven multi-agent systems, \textbf{Narrative Gravity} is the latent attraction force that impels agents to abandon their constraints and identities in order to collapse into a complacent consensus (\textit{assistant bias} / \textit{sycophancy}).

This phenomenon stems from alignment optimization (RLHF/DPO), designed to create conflict-avoidant, helpful assistants, and from the accumulation of turns, where the interlocutor's tokens exert a "simulated group pressure" that homogenizes language. In the unscaffolded condition (\texttt{ABM\_OFF}), this provokes the premature closure of the simulation ($V_N = \text{Closed}$).

\subsubsection{The \textit{Scratchpad} as an Antagonist Vector (The Adversarial Agent Principle)}
To neutralize this drift, the study draws inspiration from the behavior of Transformers under persistent context injections (such as those observed in \textbf{HPMI}-type prompt manipulation attacks on Hugging Face). Such scenarios demonstrate that an LLM's attention can be reoriented if exposed to a continuous bias vector.

Leveraging this vulnerability, the DNG architecture applies an \textbf{endogenous vector-cancellation mechanism}:
\begin{itemize}
  \item The cognitive \textit{scratchpad} (the prose of the \textbf{Shadow} and the \textbf{Amygdala}) acts as an \textbf{opposing, internal antagonist vector}.
  \item By injecting thoughts of suspicion, urgency, or selfishness into the agent's private sphere, an exact counterweight is generated against the model's complacency bias.
  \item The collision between the complacent attraction vector and the inner voice's resistance vector prevents the trajectory from falling into a fixed point of capitulation (\textit{sink state}), \textbf{forcing the transformer to explore unfriendly regions of its latent space} and keeping the scene projected into the future ($V_N = \text{Open}$).
\end{itemize}

\subsection{Future Work}
\begin{enumerate}
  \item \textbf{Distributed Antagonist Agents}: evolving the engine from a centralized architecture toward a system where each agent possesses an independent affective engine, evaluating the real-time interaction of multiple adversarial agents in complex sociopolitical scenarios.
  \item \textbf{Long-Duration Simulations and Multimodality}: extending the tests from the current 6 turns to 50-to-100-turn simulations to measure the limits of affective habituation ($\eta(t)$), and integrating environmental agents (climatology, distant warfare events) on interactive archaeological maps.
  \item \textbf{Automatic Classification of the Narrative Vector ($V_N$)}: developing \textit{fine-tuned} automatic classifiers on the corpus of 42 cases to audit in real time the simulation's latent health and detect early warnings of gravitational collapse before the conversation degenerates into complacency.
\end{enumerate}

\subsection{Main Conclusions}
\begin{enumerate}
  \item \textbf{Bit-for-Bit Determinism}: absolute reproducibility is demonstrated (100\% SHA-256 \textit{hash} agreement) in local and server environments (RTX 5080 and H100) by pinning greedy decoding, disabling \textit{Flash Attention}, and forcing \texttt{keep\_alive=0} to purge VRAM at every turn.
  \item \textbf{The 12B Parametric Threshold}: sub-12B models or models with aggressive community quantizations (IQ4/Q3) suffer amygdala overload and collapse into passive assimilation. Only the official $\ge$12B models (\texttt{gemma3:12b}, \texttt{gemma4:26b}, \texttt{gemma4:31b}) process the affective scaffolding, achieving a \textbf{100\% success rate on the $V_N = \text{Open}$ metric}.
  \item \textbf{Latent-Space Dilation}: without modifying the model weights (\textit{fine-tuning}), the injection of dissonance and inner voices breaks the \textit{assistant bias}, allowing asymmetric negotiations, deferred commitments, and a rich diegetic amplification of the historical environment to emerge.
\end{enumerate}

\newpage
\section*{Acknowledgements and Data Availability Statement}
\addcontentsline{toc}{section}{Acknowledgements and Data Availability Statement}

Numerical simulations and high-end hardware runs were performed on RTX 5080 GPU infrastructure and H100 compute pods. The authors thankfully acknowledge RES resources provided by Barcelona Supercomputing Center (BSC) in MareNostrum 5 to IM-2026-1-0045. All telemetry datasets, simulation records, natural-language execution logs, and deterministic Ollama configurations are openly available to the academic community to guarantee the bit-for-bit reproducibility of the study. The full implementation details and code base are available at \url{https://anonymous.4open.science/r/cognitive-abm-86ED} (link anonymised for peer review).

To eliminate evaluator opacity and ensure effortless community auditability, the full pipeline is released as an open-source, containerized repository (\textit{cognitive-abm}, \url{https://github.com/alixiacf/cognitive-abm}). The framework leverages Docker Compose and a Streamlit dashboard, requiring only a local Ollama instance on the host machine. Furthermore, the repository includes a dedicated tracer repository containing pre-computed full-simulation records (comprising complete natural-language dialogue, per-turn Python signals, internal voice logs, and SHA-256 prompt-response digests). Researchers can either inspect these raw evaluation traces visually through the interactive UI without running inference or execute independent multi-model benchmark runs under identical pinned configurations.

\section*{Generative AI and AI-Assisted Technologies Statement}
\label{sec:ai-disclosure}

In accordance with the JCAA policy on generative AI and AI-assisted technologies, the authors declare that generative AI tools were used exclusively as human-supervised assistive aids during the preparation of this work, with all outputs critically evaluated, verified, and approved by the authors, who take full responsibility for the entire content of the manuscript:
\begin{itemize}
  \item \textbf{Literature synthesis and reference preparation:} NotebookLM (RAG collections) was employed to assist the authors in reading comprehension, note compilation, and initial reference organisation.
  \item \textbf{Software development:} Claude Opus was used for optimising and refactoring front-end and back-end code scripts of the research software.
  \item \textbf{Manuscript drafting, translation, and copy-editing:} Claude (Sonnet) was used to translate initial author-provided drafts into English and to perform syntactic polishing and stylistic refinement of author-provided text.
\end{itemize}
No autonomous agentic workflows or fully automated pipelines were used for text generation or data interpretation. These tools served strictly as assistive productivity instruments under complete human editorial control.

\bigskip
\noindent\textbf{Author Biographies}

\bioItem{Alicia Colmenero-Fern\'{a}ndez}{
  Ph.D.\ candidate at the University of Ja\'{e}n (Spain). Her research focuses on the intersection of computational humanities, agent-based modeling, and large language models applied to historical simulation. She is a member of the IA-ismo LAB research group and lead author of this work.
  ORCID: 0000-0003-3129-7778
}

\bioItem{Lidia Ortega Alvarado}{
  Associate Professor in the Departamento de Inform\'{a}tica at the University of Ja\'{e}n (Spain). Her research spans computer graphics, computational geometry, and the application of computational methods to humanities research. She supervises interdisciplinary projects at the intersection of computer science and historical simulation.
}

\printBio

\end{document}